\documentclass[10pt,twocolumn]{IEEEtran}
\usepackage{graphicx}          
\usepackage{amsmath}
\usepackage{tabularx}
\usepackage{amsfonts}
\usepackage{url}
\usepackage{verbatim}
\usepackage{times,epsfig}
\usepackage{makecell}
\usepackage{psfrag}
\usepackage{subfigure}
\usepackage{stfloats}
\usepackage{caption}
\usepackage{makecell}
\usepackage{setspace}
\usepackage{diagbox}
\usepackage{amssymb}
\usepackage{multirow}
\usepackage{colortbl}
\usepackage[justification=centering]{caption}
\usepackage{fancyhdr}
\usepackage[table,xcdraw]{xcolor}

\begin{document}
	
	\pagenumbering{arabic}
	
	%\title{A Secure and Reliable Transfer Learning Framework for Emerging 6G Services and	Applications}
	% Quantum Key Distribution Service Provision
	% \title{Joint pretrained Foundation Model Caching and Inference in Mobile Edge Computing: A Use Case of Serving Wireless AIGC Services}
 % \title{Serving Pretrained Foundation Models in Mobile Edge Computing: A Joint Model Caching and Inference Framework for Wireless AIGC Services}
 % \fontsize{22}{24}\selectfont
 % \title{Edge Model Caching and Inference for Pretrained Foundation Models of Mobile AIGC Services in Metaverse
  \title{Sparks of GPTs in Edge Intelligence for Metaverse: Caching and Inference for Mobile AIGC Services
 % : A Joint Caching and Inference Framework
\author{Minrui Xu, Dusit Niyato, \emph{Fellow, IEEE}, Hongliang Zhang, Jiawen Kang, Zehui Xiong,\\ Shiwen Mao, \emph{Fellow, IEEE}, and Zhu Han, \emph{Fellow, IEEE}
		\thanks{Minrui~Xu and Dusit~Niyato are with the School of Computer Science and Engineering, Nanyang Technological University, Singapore 639798, Singapore. Hongliang~Zhang is with the School of Electronics, Peking University, Beijing 100871, China. Jiawen~Kang is with the School of Automation, Guangdong University of Technology, China. Zehui~Xiong is with the Pillar of Information Systems Technology and Design, Singapore University of Technology and Design, Singapore 487372, Singapore. Shiwen~Mao is with the Department of Electrical and Computer Engineering, Auburn University, Auburn, AL 36849-5201 USA. Zhu~Han is with the Department of Electrical and Computer Engineering, University of Houston, Houston, TX 77004 USA, and also with the Department of Computer Science and Engineering, Kyung Hee University, Seoul 446-701, South Korea.}
	}}
 % \title{Scalable pretrained Foundation Model Serving Systems in Mobile Edge Computing: A Use Case of Wireless AIGC Services}
	% \author{The 
	% \thanks{The authors' information.
 % }
	% }
\maketitle
\pagestyle{headings}

\begin{abstract}
Aiming at achieving artificial general intelligence (AGI) for Metaverse, pretrained foundation models (PFMs), e.g., generative pretrained transformers (GPTs), can effectively provide various AI services, such as autonomous driving, digital twins, and AI-generated content (AIGC) for extended reality. With the advantages of low latency and privacy-preserving, serving PFMs of mobile AI services in edge intelligence is a viable solution for caching and executing PFMs on edge servers with limited computing resources and GPU memory. However, PFMs typically consist of billions of parameters that are computation and memory-intensive for edge servers during loading and execution. In this article, we investigate edge PFM serving problems for mobile AIGC services of Metaverse. First, we introduce the fundamentals of PFMs and discuss their characteristic fine-tuning and inference methods in edge intelligence. Then, we propose a novel framework of joint model caching and inference for managing models and allocating resources to satisfy users' requests efficiently. Furthermore, considering the in-context learning ability of PFMs, we propose a new metric to evaluate the freshness and relevance between examples in demonstrations and executing tasks, namely the Age of Context (AoC). Finally, we propose a least context algorithm for managing cached models at edge servers by balancing the tradeoff among latency, energy consumption, and accuracy. 

% Foundation models 具有涌现的能力 能够用世界知识同时解决分类,决策和生成任务.
% 在移动边缘计算中, 提供基于foundation models 的AIGC服务
\end{abstract}

\section{Introduction}

Towards artificial general intelligence (AGI) in Metaverse~\cite{bubeck2023sparks, zhou2022vetaverse}, pretrained foundation models (PFMs), e.g., generative pretrained transformers (GPTs)~\cite{brown2020language}, with billions of parameters achieve great success across a variety of fields over the past few years due to their effectiveness at demonstrating emergence abilities in downstream tasks with different data modalities~\cite{zhou2023comprehensive}. The pretraining approach offers a reasonable parameter initialization for extensive downstream applications, such as object detection, image generation, and text retrieval. Therefore, PFMs, including language foundation models (LFMs), visual foundation models (VFMs), and multimodal foundation models (MFMs), are in the paradigm of transfer learning that can generalize to new tasks and domains without any task-specific data during pretraining. 

PFMs can empower a multitude of intelligent services for Metaverse, such as autonomous driving, digital twins (DTs), and  artificial intelligence-generated content (AIGC) for extended reality (XR). For instance, PFMs can facilitate complex driving decisions and generate traffic simulations for autonomous driving~\cite{xu2023generative}. Moreover, PFMs can help understand and respond to human emotions and behaviors during immersive human-avatar interactions. For example, based on the GPT-3~\cite{brown2020language}, which is an LFM with 175 billion parameters, ChatGPT\footnote{https://openai.com/blog/chatgpt/} enables long and fluent conversations with humans using world knowledge and contextual awareness. In addition to serving PFMs at cloud servers, edge servers equipped with GPU resources can also support fine-tuning and inference processes of Metaverse services, which brings the sparks of GPTs to mobile edge networks.
% As PFMs are typically large and require intensive computing power and GPU memory to execute, 
Therefore, deploying PFMs in mobile edge networks allows the provision of low-latency, personalized, customized, and privacy-preserving AI services.
% Furthermore, PFMs-based analytical AI services, decision AI services, and generative AI services can provide process optimization and efficiency improvements in a wide range of mechanized and creative work.

However, compared to cloud servers, resource-constraint edge servers cannot load all PFMs simultaneously to satisfy the requests of services in Metaverse. Aiming at provisioning mobile AI services in edge networks, existing works primarily focus on offloading AI services to cloud servers for remote execution or caching inference outputs at edge servers for low-latency access~\cite{gilman2019challenges}. On the one hand, offloading PFMs of AI services to cloud servers
% , which load most of PFMs and provide instant inference results of AIGC services, 
results in extra core networking latency, traffic overhead, and privacy risks for users utilizing AI services. On the other hand, caching inference outputs at edge servers is no longer efficient for provisioning real-time AI services. Therefore, directly deploying PFMs at edge servers requires effective and fine-grained resource and request management for executing AIGC requests with available computing and energy resources at edge servers.

% 第二段: 为什么要有Joint Model Caching 和 Inference
% 第三段: 实现这个有什么难度
Specifically, in contrast to existing works on joint service caching and task offloading~\cite{xu2018joint}, there are several unique difficulties for joint PFM caching and inference to balance the tradeoff among accuracy, latency, and energy consumption in edge intelligence as follows~\cite{zhou2019edge}.
\begin{itemize}
    % \item Caching-depended Computation and Energy Consumption: After models are cached, edge servers allocate computing and energy resources to execute these cached models to respond to the service requests.
    \item \textbf{Dynamic Runtime Configuration:} During the execution of PFMs, there are varying numbers of requests and performance requirements for downstream tasks, such as accuracy and latency~\cite{gilman2019challenges}. % Because of the functional equivalence of the models, edge servers can use higher-end models to handle requests for lower-end models with slightly higher resource consumption.
    \item \textbf{Equivalent Model Adaptation:} Different PFMs that can be applied to similar downstream tasks in various Metaverse services adaptively~\cite{zhou2023comprehensive}. 
    % For example, Stable Diffuiosn and DALL-E can both perform image generation tasks but these two models have different service qualities and resource consumption~\cite{rombach2022high}.
    This introduces a challenge for edge servers, as cached PFMs can be used for inference interchangeably to minimize model misses.
    \item \textbf{Continuous In-context Learning:} PFMs, like GPT-3, can continuously learn and adapt to new domains and tasks based on interactive demonstrations for personalization and customization~\cite{dong2022survey}. The ability of in-context learning enables cached PFMs to improve their performance during inference without parameter updates. This adds complexity in making cache replacement and deployment decisions, as it presents a new tradeoff among inference latency, resource consumption, and accuracy.
\end{itemize}
\begin{figure*}[t]
    \centering
    \includegraphics[width=1\linewidth]{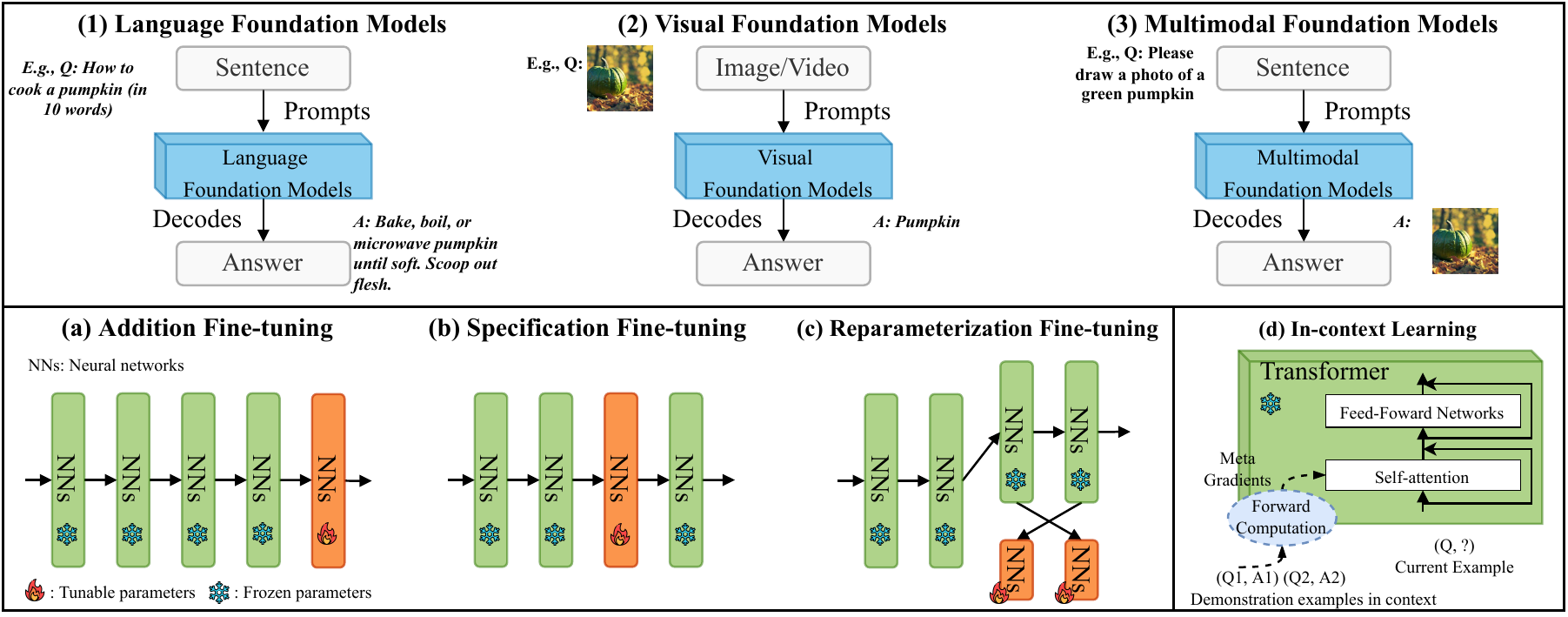}
    \caption{Categories of PFMs and their characteristic fine-tuning and inference methods. (1)-(3) The workflows of LFMs, VFMs, and MFMs. (a)-(c) The illustration of parameter-efficient fine-tuning. (d) An example of in-context learning.}
    \label{fig:pfms}
\end{figure*}
% 第四段: 

To address these issues, this article investigates the potential but scarcely studied problems of PFM caching and inference in mobile edge networks. We first introduce the fundamentals of PFMs for serving mobile AIGC services of Metaverse, and their fine-tuning, and inference methods in edge networks. Then, we present a joint model caching and inference framework in edge networks to serve PFMs of mobile AI services of Metaverse. Furthermore, we discuss potential applications and challenges of serving PFMs for Metaverse services. Finally, to balance the tradeoff among inference latency, resource consumption, and accuracy, we propose a novel metric to indicate the freshness and relevance of examples in demonstrations and current tasks, namely the \textbf{Age of Context} (AoC). The AoC follows the non-increasing utility function that affects the effective examples in context from the entirety of demonstrations resulting from historical interactions. Based on this metric and the number of examples in context, we propose a least context (LC) algorithm to manage PFMs at edge servers. Experimental results demonstrate that the proposed LC algorithm can reduce the total system cost by improving the accuracy of edge-cached PFMs, reducing offloading latency, and utilizing the caching and computing resources of edge servers efficiently.
% Finally, we study the use case of providing wireless AIGC services in MEC. Unlike existing work that utilizes online learning algorithms to predict users' requests and operate cache replacement, we develop a diffusion model-based offline learning algorithm to improve data efficiency and increase the scalability of the proposed system.
\section{Serving PFMs of Services in Metaverse}\label{sec:foundation}

% In this section, we first introduce the fundamentals of PFMs. Then, we highlight the differences in fine-tuning and inference of PFMs, respectively.

\subsection{Fundamentals of Pretrained Foundation Models}

PFMs belong to the transfer learning paradigm that is used to initialize parameters for downstream tasks. PFMs, such as BERT, GPT-3, Stable Diffusion, CLIP, and ChatGPT, leverage large-scale datasets and pretraining techniques to provide reasonable parameter initialization for various AI services~\cite{zhou2023comprehensive}. 
% Pretraining involves training models on data before applying them to the task at hand; this helps reduce overfitting by providing better generalization capabilities than if the model was trained from scratch. 
As shown in Fig.~\ref{fig:pfms}, there are primarily three types of PFMs, i.e., LFMs, VFMs, and MFMs, which are widely employed to provide AI services.

\subsubsection{Language Foundation Models}
LFMs, also known as large-scale language models, are PFMs designed to understand, process, and generate human languages. LFMs are trained on massive amounts of text data and can develop a broad understanding of language, including grammar, syntax, semantics, and even some aspects of common knowledge. Two examples of PFMs are GPT and ChatGPT, which have demonstrated impressive abilities in natural language understanding and generation. GPT-3 can enable conversations with humans based on world knowledge and contextual awareness, while ChatGPT is designed to generate human-like responses in a chatbot setting. These models employ self-attention mechanisms to better understand the context and relationships between words in a given text and can be adopted in various downstream tasks, such as sentiment analysis, machine translation, text summarization, question-answering, and text generation. 
% While these models have improved the field of NLP, they also raise concerns regarding ethical issues such as biases present in the training data and potential misuse of the technology.

\subsubsection{Visual Foundation Models}

VFMs specialize in understanding and generating complex images and videos, which are designed to process visual information and generate target outputs. VFMs have shown great potential in advancing the field of computer vision, but they are computing-intensive, particularly during the inference stage. For example, the U-Net in Stable Diffusion~\cite{rombach2022high}, which is a generative model that can produce high-quality images by iteratively refining a noise vector. Stable Diffusion uses a diffusion process to create realistic and high-quality images, and it has been shown to outperform other generative models on a variety of tasks. 
% While the computational cost of training and inference for Stable Diffusion can be high~\cite{rombach2022high}, its ability to generate high-quality images makes it a valuable tool for a wide range of applications. 
% Another example is Midjourney's diffusion-based, text-to-image service, which has more than 4.4 million users and requires more than 10,000 NVIDIA GPUs mainly for AI inference. While the computational cost of inference for Midjourney's service is significant, it demonstrates the potential for VFMs to enable new and innovative applications in the field of computer vision.
\subsubsection{Multimodal Foundation Models}

MFMs can process multiple types of data, such as text, images, and audio simultaneously. They are trained on datasets containing various data modalities to learn the relationships, patterns, and structures within and across different data types. For instance, CLIP is one of the MFMs that classify images based on textual descriptions~\cite{cherti2022reproducible}, which uses contrastive learning to train on text and image pairs, distinguishing between positive and negative pairs. During inference, the model takes in an image and a textual description and outputs a score representing the likelihood that the image matches the description, calculated through a dot product. Furthermore, MFMs can be fine-tuned on specific tasks by training them on a smaller dataset.

\subsection{Fine-Tuning of Pretrained Foundation Models}

Fine-tuning refers to the process of improving the performance of PFMs to a specific downstream task by updating its parameters. Since PFMs usually consist of billions of parameters, the fine-tuning process is computationally intensive. Therefore, parameter-efficient fine-tuning of PFMs is utilized for achieving comparable performance to traditional fine-tuning while reducing resource consumption~\cite{xu2023unleashing}. As shown in Fig.~\ref{fig:pfms}, parameter-efficient fine-tuning can be categorized into three types, including addition-based, specification-based, and reparameterization-based methods as follows~\cite{ding2023parameter}. 
\begin{figure*}
    \centering
    \includegraphics[width=0.8\linewidth]{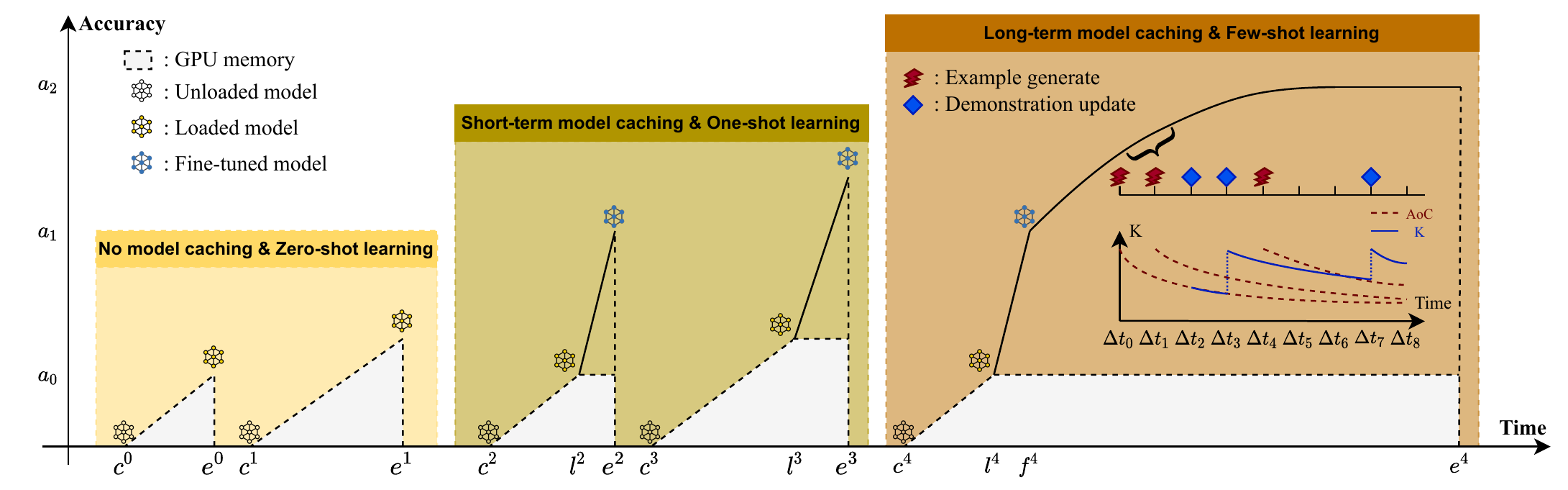}
    \caption{An illustration of the performance of zero-, one-, and few-shot accuracy under different model caching settings~\cite{brown2020language}.}
    \label{fig:accuracy}
\end{figure*}
\begin{itemize}
    \item Addition-based methods involve adding a small number of parameters to the PFMs and fine-tuning them. These methods, which include scalar addition, vector addition, and layer addition, add parameters to the PFMs that are specific to the fine-tuning data. For instance, such parameters include additional layers or heads after the output layer of PFMs.
    \item Specification-based methods modify the architecture of PFMs to better suit downstream tasks. These methods, such as layer removal, layer replacement, and layer scaling, adjust the PFMs' parameters and architecture to improve performance.

    \item Reparameterization-based methods reduce the number of tunable parameters in PFMs by reparameterizing their parameters. These methods, such as low-rank factorization, matrix decomposition, and subspace projection, reparameterize the PFMs to reduce the number of tunable parameters while preserving the PFMs' expressiveness.
\end{itemize}

Depending on applications such as Metaverse, the fine-tuning methods can be selected adaptively depending on the resource and performance requirements.

% As pretrained foundation models usually consist of billions of parameters, it is infeasible to fine-tune all the parameters 
\subsection{Inference of Pretrained Foundation Models}

Different from fine-tuning that updates the parameters of PFMs, the inference is to make predictions on input service requests without changing the parameters. Instead of injecting or updating neural modules in AI models, PFMs can provide accurate output for the task that does not exist in the training, fine-tuning, and inference from instructions and demonstrations from interaction without parameter updates. As shown in Fig.~\ref{fig:accuracy}, there are three scenarios during the inference of PFMs~\cite{brown2020language}, including zero-shot, one-shot, and few-shot learning. First, zero-shot learning refers to the PFMs that are evaluated on a task for which it has not been explicitly trained. Then, one-shot learning indicates the PFMs need to perform the inference for a new task based on only one example of that task. Finally, few-shot learning implies that a few demonstrations are provided before the inference of the new task. Based on the few-shot learning, the PFMs can perform a meta-gradient in the self-attention layer for adaptation to the new task. Different from fine-tuning, few-shot learning or in-context learning can perform meta-gradient in the attention layers during inference without changing its model parameters. Therefore, few-shot learning can improve the model performance based on examples in instructions and/or demonstrations. However, extra computation consumption and latency are required by processing the examples which depend on the size of the context window in PFMs. 
% \begin{itemize}
%     \item Zero-shot refers to the PFMs that are evaluated on a task for which it has not been explicitly trained, and without any additional training data. The model is expected to generalize to this new task based on its prior training on related tasks.
%     \item One-shot refers to the scenario where a model is trained on a new task with only one example of that task. The model is expected to learn from this single example and generalize to new examples of the same task.
%     \item Few-shot refers to the scenario where a model is trained on a new task with only a few examples of that task. The model is expected to learn from these few examples and generalize to new examples of the same task.
% \end{itemize}
% Although these methods are named after learning, they do not actually update the model parameters. The term ``learning" more accurately refers to the fact that they gain new knowledge during the inference process and can learn new tasks.

\section{Joint Model Caching and Inference Framework}\label{sec:caching}
To serve PFMs in edge intelligence for Metaverse, we develop a framework of joint model caching and inference to satisfy service level objectives by utilizing caching, computing, and communication resources in mobile edge networks.
Unlike content caching in content delivery networks (CDNs), such as text, images, and videos, the cached models have different cache structures. The cache structure in CDNs is static, with fixed cache sizes and independent of computation resources~\cite{xu2018joint}. However, due to the flexible configuration of PFMs, the cache structures are dynamic, adjusting to the service requirements in the Metaverse service layer and depending on computation resources during fine-tuning and inference. In the framework, we discuss the model caching configuration and model caching and eviction policy in the PFM layer. Then, we introduce a collaborative mobile edge-cloud layer for joint model caching and inference.

\subsection{Model Caching Configuration}

The configuration of each cached PFM consists of the following information.
% \begin{figure}
%     \centering
%     \includegraphics[width=1\linewidth]{Figs/image.png}
%     \caption{Runtime GPU memory and inference latency of Stable Diffusion under different batch sizes during inference.}
%     \label{fig:sd}
% \end{figure}
\begin{itemize}
    \item \textbf{Frequency of Use:} Frequency of use for PFMs refers to the rate at which a particular model is executed for services in Metaverse. It can be measured in terms of the number of requests per second, the total time spent on processing each request, and other metrics that measure how often a PFM is being utilized.
    \item \textbf{Model Sizes:} Model size indicates the number of parameters, weights, and other necessary components of PFMs, which affects the latency and energy cost of edge servers for loading and executing PFMs~\cite{gilman2019challenges}. 
    % Then, the model size is closely related to runtime GPU memory after the PFMs are loaded into the servers.
    \item \textbf{Runtime GPU Memory:} Runtime GPU memory measures how much RAM or VRAM (video random access memory) is needed by loading the PFM to execute on a given edge/cloud server with its current configuration settings. The runtime GPU memory not only depends on the model sizes but also the runtime precision configuration of the precision. 
    % In general, models with higher precision require more memory to store their weights and activations, which can lead to higher GPU memory usage. However, using lower-precision models can lead to a decrease in model accuracy. 
    Therefore, there is a trade-off between model precision and GPU memory usage.
    \item \textbf{Model Speed:} Model speed of PFMs refers to the time complexity or speed at which a particular model can process inference requests. It is usually measured in terms of inference times, i.e., how long it takes for a PFM to complete its task given certain inputs and parameters. Model Speed also has implications on accuracy as faster processing often leads to lower accuracy due to less computation power being available for each request.
    \item \textbf{Model Accuracy:} Model accuracy of PFMs refers to the degree of correctness or precision with which a model can predict outcomes. 
    % It is usually measured in terms of accuracy scores, i.e., how close a PFM's predictions are compared to actual results given certain inputs and parameters. 
    For PFMs, model accuracy has implications on speed as higher accuracy often requires more computation power for each request, leading to longer processing times overall.
    \item \textbf{Number of Examples in Context:} Cached PFMs can accumulate instructions and demonstrations while processing inference requests. The number of examples in context represents the number of related examples in demonstrations the PFMs have gathered. Due to the in-context learning ability of PFMs, the number of examples in context can also impact the accuracy of the models. The size of the context window limits the maximum number of examples in context that can be utilized for each PFM during each inference.
\end{itemize}
\subsection{Model Caching and Eviction}
Since the cache structure of PFMs is more complicated than traditional web/content caching, the model caching and eviction are also more intractable. Model caching and eviction mainly consist of two types of operations, i.e., the passive and active operation as well as binary and partial operation.
\begin{itemize}
    \item \textbf{Passive and Active Caching and Eviction:} Passive caching is a reactive approach where models are evicted from the cache only when there is not enough GPU memory to load a requested model. Additionally, active caching is a proactive approach where models are evicted and loaded into GPU memory based on predictions of future demand. Active caching can be more efficient than passive caching~\cite{gilman2019challenges}, but requires more sophisticated prediction algorithms and can be less responsive to sudden changes in demand.
    \item \textbf{Binary and Partial Caching and Eviction:} Binary caching involves loading the entire model into GPU memory before starting inference. In contrast, with partial caching, only a portion of the model is loaded into memory, and inference can begin using that portion. This approach provides a lower level of inference but can be useful when memory resources are limited. When additional memory becomes available, the remaining portions of the model can be loaded into memory, improving inference quality.
\end{itemize}
If the framework can accurately predict future service request demand, it can leverage active and partial caching to enhance the quality of mobile AI services in Metaverse and reduce resource consumption in mobile edge networks.

\subsection{Collaborative Mobile Edge-Cloud Caching and Inference}
\begin{figure}[t]
    \centering
    \includegraphics[width=1\linewidth]{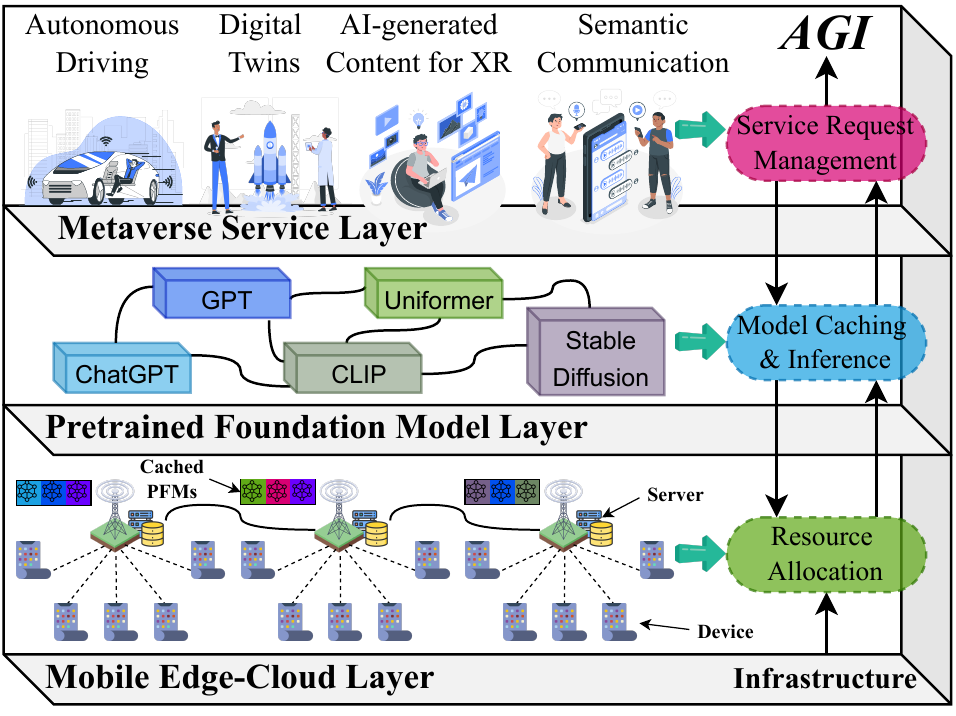}
    \caption{An illustration of the collaborative mobile edge-cloud computing architecture for serving PFMs for Metaverse.}
    \label{fig:layer}
\end{figure}
As shown in Fig.~\ref{fig:layer}, collaborative resource allocation among heterogeneous mobile edge-cloud infrastructures is critical in paving the way toward AGI at the edge.
\subsubsection{Mobile Caching and Inference} Pedestrians and vehicles can process the services to cache and execute PFMs with their local computing resources on mobile devices or devices nearby. This solution can be useful in situations where internet connectivity is limited or unreliable. %, as it allows users to perform inference tasks without relying on a centralized cloud infrastructure. 
% Additionally, mobile caching and inference can improve privacy and security by allowing users to keep their data and models on their own devices, rather than transmitting them to a remote server. 
\subsubsection{Edge Caching and Inference} When the local resources of mobile devices and vehicles are not enough for executing PFMs, offloading these services to edge servers via radio access networks become an alternative solution for enabling AI services on edge servers with limited resources. However, due to the limited GPU resources of edge servers, they can only cache several PFMs to react to the user's request. If the edge server does not cache the model requested by the user, it can migrate the user's request to the cloud for execution via core networks or load the model and then execute the model requested by the user. This approach can improve response time and reduce the load on the cloud infrastructure, making it more scalable and cost-effective.
\subsubsection{Cloud Caching and Inference} Cloud caching and inference solutions involve the utilization of powerful cloud servers to provide almost all PFMs for serving users' requests. However, offloading services to cloud servers incur additional core network latency, which might cause congestion in core networks if there are too many service requests. 
% Additionally, user privacy cannot be guaranteed when utilizing cloud caching inference, as user requests may need to be transmitted to a remote server. Despite these challenges, cloud caching inference can be a useful approach for enabling AI services that require complex or resource-intensive models.

\subsection{Model Caching and Eviction Policy}

To design the model caching and eviction policy, three issues should be considered carefully.
\begin{itemize}
\item \textbf{Reducing model miss rate:} Actively preloading models and optimizing GPU utilization through dynamic scheduling of AI models can minimize latency and model miss rates, streamlining memory use and request handling.
\item \textbf{Addressing model misses:} Handling model misses at edge servers involves offloading service requests to cloud servers, incurring extra core network latency, or loading missing models, leading to switching costs, such as additional latency and energy consumption for allocating resources as well as hardware wear-and-tear.
\item \textbf{Timing model cache decisions:} Making cache decisions when the model is first loaded and upon receiving new requests enables dynamic adjustments based on current conditions and usage patterns, promoting efficient caching and lower latency responses.
\end{itemize}
% \begin{itemize}
%     \item How to reduce the model miss rate: To minimize the model missing rate, models can be actively preloaded into memory before they are requested so that the time taken to load them is minimized. In addition, optimizing GPU utilization through dynamic scheduling of requests in order to minimize latency caused due to waiting periods between requests.
%     \item How to handle a model miss: There are two methods to handle a model miss at edge servers, i.e., offloading service requests to cloud servers or loading missing models. To offload requests to edge servers, extra core network latency is a cost to the users. In the meanwhile, loading missing models can also lead to extra model loading latency.
%     \item When to make model cache decisions: Model cache decisions should be made when the model is first loaded into memory, as well as whenever a new request for that particular model comes in. This allows for dynamic adjustments to be made based on current conditions and usage patterns of the system. Additionally, it also helps ensure that models are cached efficiently so they can quickly respond to requests with lower latency.
% \end{itemize}

Therefore, an effective and efficient caching algorithm in this framework should address these three questions properly. Then, we can summarize two remarks for serving PFMs of mobile AI services as follows.

% How to schedule Inference request

% How to reduce the model miss rate

% How to handle a model miss

% When to make model cache decisions

\textbf{Remark 1:} Different from traditional edge content caching in content delivery networks, whose cache structures are static and independent from computation, the cache structures can be dynamic based on the service runtime configuration, such as batch size. This makes the cache loading and eviction of AI services more complex, which requires not only the consideration of user preferences but also the prediction of the intensity of future service requests.

\textbf{Remark 2:} Unlike traditional computation and task offloading in mobile edge networks, where different computation tasks are independent, the inference tasks of PFM-related services are in-contextual. Therefore, before performing these inference tasks, the AIGC model needs to be preloaded into the edge servers' GPU memory, which can cache a limited number of models to provide AI services. 
% If the model is not cached in the GPU memory when the AIGC service request is received, the edge server can either load the model and process it or offload it directly to the cloud data center for processing. 
Furthermore, as more in-context examples are collected during the interaction, the performance of cached services can be further improved~\cite{brown2020language}.

\section{Potential Applications and Challenges}\label{sec:application}

\subsection{Applications}

The potential PFM-based applications in Metaverse include autonomous driving, DTs, semantic communication, and AIGC for XR.

\subsubsection{Autonomous Driving} Autonomous driving in Metaverse necessitates AI services such as traffic and driving simulation, which are dependent on computationally-intensive PFMs~\cite{xu2023generative}. To enable this on resource-limited edge servers, model caching, and efficient inference scheduling are essential. In autonomous driving, active model switching can enhance traffic efficiency and safety by adapting to changing road conditions or traffic patterns. 
% Employing functionally-equivalent models with distinct architectures and computational demands, autonomous driving systems can optimize model utilization while reducing the computational burden on vehicles and roadside units, thus ensuring safer and more efficient decision-making.

\subsubsection{Digital Twins} DTs, virtual representations of physical objects or systems, utilize PFMs for AI capabilities like predictive maintenance, anomaly detection, and optimization. The complexity of these systems demands numerous PFMs for accurate modeling across diverse scenarios. Therefore, effective management of the vast number of PFMs, maintaining quality and consistency, and optimizing caching and inference policies can enable DTs to accurately represent physical systems, thus enhancing decision-making and operational efficiency.

\subsubsection{Semantic Communication} Semantic Communication, a novel paradigm that employs semantic representations, can transform wireless communication systems' design and operation. Its device-to-device pattern enables efficient and secure communication without centralized cloud infrastructure. However, this pattern necessitates advanced model caching algorithms to manage edge servers' limited resources while ensuring cached models' quality and consistency. Implementing progressive caching techniques like active and partial caching can optimize the device-to-device pattern, leading to faster and more reliable AI services on edge servers.

% Semantic Communication is a promising paradigm that can revolutionize the design and operation of wireless communication systems. It involves the use of AI to create semantic representations that imbue meaning, significance, and structure into every information transfer over a wireless network. One of the key advantages of Semantic Communication is the device-to-device pattern~\cite{chaccour2022less}, which allows for more efficient and secure communication without the need for centralized cloud infrastructure. However, this pattern also requires active and efficient model caching algorithms to manage the limited resources of edge devices and ensure the quality and consistency of cached models. With the development of advanced caching techniques, such as active caching and partial caching, the device-to-device pattern of Semantic Communication can be further optimized to deliver faster and more reliable AI services on edge devices.

\subsubsection{AIGC for XR} AIGC is generated by AI methods that utilize PFMs to create content that resembles human-produced content~\cite{xu2023unleashing}. To provide AI-generated XR services, multiple PFMs are integrated to handle different types of data and produce relevant and meaningful 3D immersive content. The model caching algorithm ensures that the PFMs work smoothly together, maintaining seamless and immersive experiences for Metaverse users. Achieving this requires careful consideration of the interplay between PFMs and the development of advanced context-aware caching algorithms for efficient cached model management and coordination.

% \usepackage{multirow}
% \usepackage{multirow}
% \usepackage{multirow}
\begin{table*}[t]
\small\centering
\caption{Detailed parameters and performance of PFMs (K=number of examples in context).}
\label{tab:my-table}
\begin{tabular}{|c|c|c|c|c|c|ccc|}
\hline
 &
   &
   &
   &
   &
   &
  \multicolumn{3}{c|}{\textbf{Model Performance Score}} \\ \cline{7-9} 
\multirow{-2}{*}{} &
  \multirow{-2}{*}{\textbf{Models}} &
  \multirow{-2}{*}{\textbf{Downstream Tasks}} &
  \multirow{-2}{*}{\textbf{Model Size (M)}} &
  \multirow{-2}{*}{\textbf{GFLOPs}} &
  \multirow{-2}{*}{\textbf{K}} &
  \multicolumn{1}{c|}{Zero-shot} &
  \multicolumn{1}{c|}{One-shot} &
  Few-shot\\ \hline
 &
   &
  Translation &
   &
   &
  64 &
  \multicolumn{1}{c|}{15.45} &
  \multicolumn{1}{c|}{26.12} &
  30.83 \\ \cline{3-3} \cline{6-9} 
 &
   &
  Basic arithmetic &
   &
   &
  50 &
  \multicolumn{1}{c|}{3.79} &
  \multicolumn{1}{c|}{15.98} &
  14.34 \\ \cline{3-3} \cline{6-9} 
 &
  \multirow{-3}{*}{GPT3-13B~\cite{brown2020language}} &
  SuperGLUE &
  \multirow{-3}{*}{12850} &
  \multirow{-3}{*}{26.54} &
  32 &
  \multicolumn{1}{c|}{54.4} &
  \multicolumn{1}{c|}{64.3} &
  66.9 \\ \cline{2-9} 
 &
    &
   Translation &
    &
    &
   64 &
  \multicolumn{1}{c|}{ 22.03} &
  \multicolumn{1}{c|}{ 29.63} &
   33.77 \\ \cline{3-3} \cline{6-9} 
 &
    &
   Basic arithmetic &
    &
    &
   50 &
  \multicolumn{1}{c|}{ 25.99} &
  \multicolumn{1}{c|}{ 40.71} &
   49.55 \\ \cline{3-3} \cline{6-9} 
\multirow{-6}{*}{LFMs} &
  \multirow{-3}{*}{ GPT-3-175B~\cite{brown2020language}} &
   SuperGLUE &
  \multirow{-3}{*}{ 174600} &
  \multirow{-3}{*}{ 354.03} &
   32 &
  \multicolumn{1}{c|}{ 58.2} &
  \multicolumn{1}{c|}{ 68.9} &
   73.2 \\ \hline
 &
   &
  Image classification &
  22 &
  3.6 &
   &
  \multicolumn{1}{c|}{82.9} &
  \multicolumn{1}{c|}{-} &
  - \\ \cline{3-5} \cline{7-9} 
 &
   &
  Video classification &
  22 &
  167 &
   &
  \multicolumn{1}{c|}{82.8} &
  \multicolumn{1}{c|}{-} &
  - \\ \cline{3-5} \cline{7-9} 
 &
   &
  \begin{tabular}[c]{@{}c@{}}Object detection and \\ instance segmentation\end{tabular} &
  41 &
  269 &
   &
  \multicolumn{1}{c|}{45.6} &
  \multicolumn{1}{c|}{-} &
  - \\ \cline{3-5} \cline{7-9} 
 &
   &
  Semantic segmentation &
  25 &
  247 &
   &
  \multicolumn{1}{c|}{46.6} &
  \multicolumn{1}{c|}{-} &
  - \\ \cline{3-5} \cline{7-9} 
 &
  \multirow{-5}{*}{UniFormer-S~\cite{li2022uniformer}} &
  Pose estimation &
  25 &
  4.7 &
  \multirow{-5}{*}{-} &
  \multicolumn{1}{c|}{74.0} &
  \multicolumn{1}{c|}{-} &
  - \\ \cline{2-9} 
 &
    &
   Image classification &
   50 &
   8.3 &
    &
  \multicolumn{1}{c|}{ 83.9} &
  \multicolumn{1}{c|}{ -} &
   - \\ \cline{3-5} \cline{7-9} 
 &
    &
   Video classification &
   22&
   389 &
    &
  \multicolumn{1}{c|}{ 84.0} &
  \multicolumn{1}{c|}{ -} &
   - \\ \cline{3-5} \cline{7-9} 
 &
    &
   \begin{tabular}[c]{@{}c@{}}Object detection and \\ instance segmentation\end{tabular} &
   69&
   399 &
    &
  \multicolumn{1}{c|}{ 47.4} &
  \multicolumn{1}{c|}{ -} &
   - \\ \cline{3-5} \cline{7-9} 
 &
    &
   Semantic segmentation &
   54&
   471 &
    &
  \multicolumn{1}{c|}{ 48.0} &
  \multicolumn{1}{c|}{ -} &
   - \\ \cline{3-5} \cline{7-9} 
\multirow{-10}{*}{VFMs} &
  \multirow{-5}{*}{ UniFormer-B~\cite{li2022uniformer}} &
   Pose estimation &
   54&
   9.2 &
  \multirow{-5}{*}{ -} &
  \multicolumn{1}{c|}{ 75.0} &
  \multicolumn{1}{c|}{ -} &
   - \\ \hline
 &
   &
  Classification &
   &
   &
   &
  \multicolumn{1}{c|}{75.20} &
  \multicolumn{1}{c|}{-} &
  - \\ \cline{3-3} \cline{7-9} 
 &
   &
  Image retrieval &
   &
   &
   &
  \multicolumn{1}{c|}{71.08} &
  \multicolumn{1}{c|}{-} &
  - \\ \cline{3-3} \cline{7-9} 
 &
  \multirow{-3}{*}{CLIP-ViT-L/14~\cite{cherti2022reproducible}} &
  Text retrieval &
  \multirow{-3}{*}{428} &
  \multirow{-3}{*}{175.5} &
  \multirow{-3}{*}{-} &
  \multicolumn{1}{c|}{84.00} &
  \multicolumn{1}{c|}{-} &
  - \\ \cline{2-9} 
 &
    &
   Classification &
    &
    &
    &
  \multicolumn{1}{c|}{ 77.97} &
  \multicolumn{1}{c|}{ -} &
   - \\ \cline{3-3} \cline{7-9} 
 &
    &
   Image retrieval &
    &
    &
    &
  \multicolumn{1}{c|}{ 73.43} &
  \multicolumn{1}{c|}{ -} &
   - \\ \cline{3-3} \cline{7-9} 
\multirow{-6}{*}{MFMs} &
  \multirow{-3}{*}{ CLIP-ViT-H/14~\cite{cherti2022reproducible}} &
   Text retrieval &
  \multirow{-3}{*}{ 986} &
  \multirow{-3}{*}{ 381.9} &
  \multirow{-3}{*}{ -} &
  \multicolumn{1}{c|}{ 86.04} &
  \multicolumn{1}{c|}{ -} &
   - \\ \hline
\end{tabular}
\end{table*}

\subsection{Challenges}
% To serve PFMs of mobile AIGC services in Metaverse, there are several critical challenges in the framework that need to be addressed.
\subsubsection{Dynamic User Service Requests and Objectives} 

Joint caching and inference services at edge servers face challenges due to dynamic user requests and objectives, such as service latency and accuracy. To tackle these challenges, edge servers must efficiently manage limited resources, ensure cached model quality and consistency, and design joint caching and inference policies to satisfy users' objectives, considering factors like model size, frequency of use, and accuracy. Addressing these challenges might require edge servers to develop prediction models for various mobile AI services and propose active caching and inference algorithms.

\subsubsection{Heterogeneous Model Configuration and Computing Resources} Heterogeneous model configuration and computing resources present challenges in proposing joint model caching and inference algorithms. In detail, PFMs' structure and available edge servers result in varying GPU memory and compute resource requirements, which is typically formulated as an NP-hard mixed integer programming problem, complicating the optimization of caching and inference policies. Moreover, distinct model architectures and computation requirements add complexity. To address these challenges, edge servers must efficiently allocate resources according to each model's specific requirements while considering local computing device availability and capabilities.

\subsubsection{Context-aware Caching and Inference Algorithms} Co-designing caching and inference algorithms considering contextual information in mobile AI services at edge servers is challenging due to the indirect correlation between model caching and inference duration. Joint policies need to optimize resource allocation according to each model and inference requests' specific requirements while considering user objectives, model size, usage frequency, and accuracy. By co-designing caching and inference algorithms considering the number of examples in context, as shown in Fig.~\ref{fig:context}, edge servers can utilize extra computation resources for improving the accuracy of PFMs.

\section{Use Case of Serving GPTs in Edge Intelligence for Metaverse}\label{sec:usecase}

% In this section, we demonstrate the workflow of the proposed framework by providing a use case.

\subsection{Mobile AIGC Service Serving Model}

We consider an intelligent transportation system in Metaverse system with a remote cloud center, an edge server, and multiple vehicles, serving different Metaverse services, including autonomous driving, DTs, and AIGC-based XR, based on various PFMs. For instance, pedestrians and passengers can immerse themselves in Metaverse with XR by creating and interacting with AI-generated XR content synthesized by PFMs.
% The AIGC services of vehicles can be served by their mobile devices locally.
When users do not have enough resources on their devices and onboard units for executing PFMs, they need to offload requests to edge servers or cloud servers for remote execution. Usually, an AIGC service requires multiple PFMs to work in synergy to satisfy the user's requirements in Metaverse. For example, the Stable Diffusion services consist of three types of PFMs~\cite{rombach2022high}, including a Variational Autoencoder that compresses images into a smaller dimensional latent space, a pretrained CLIP ViT-L/14 for conditioning, and a U-Net block that denoises the output from forward diffusion backward to obtain a latent representation. 

% The VAE decoder generates the final image by converting the representation back into pixel space. The denoising step can be conditioned on text, images, or other modalities, and uses a cross-attention mechanism to expose the encoded conditioning data to the U-Nets. Finally, a trained CLIP ViT-L/14 text encoder is used for text conditioning. As previously mentioned, PFMs utilize meta-gradients to learn from context and improve performance as users interact with them. Therefore, contextual information has a corresponding impact on the quality of service provided by AIGC, such as the accuracy of PFMs. The size of the context window also affects the accuracy of the model, and the context's relevance decreases over time until it is no longer pertinent to the current task at hand.

\begin{figure}[t]
\vspace{-0.3cm}
    \centering
    \includegraphics[width=0.8\linewidth]{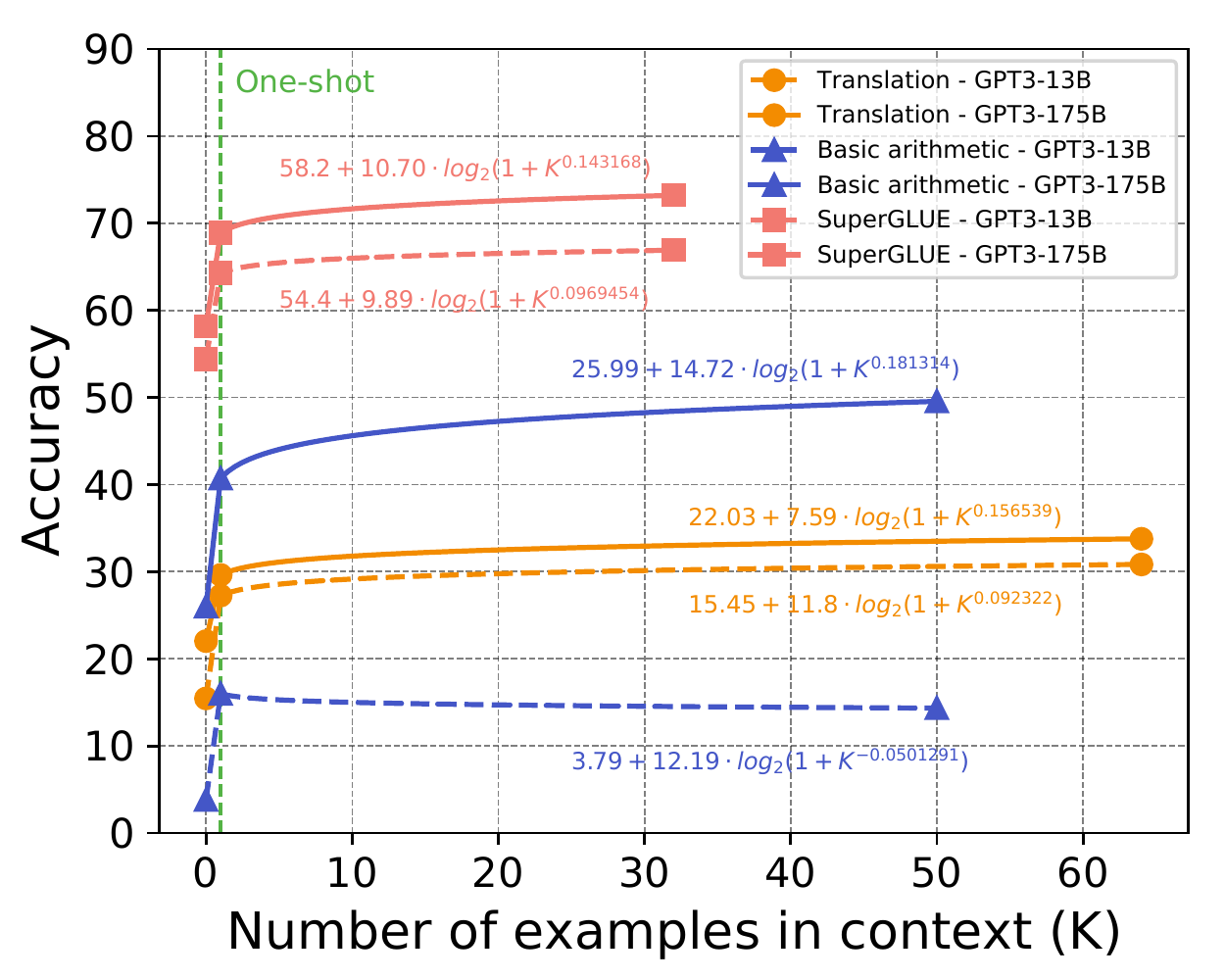}
    \caption{The accuracy in downstream tasks of GPT3-13B/ 175B versus number of examples in context. The few-shot accuracy $a_2 = a_0 + a_1 \log_2(1+K^\alpha)$, where $a_0$ is zero-shot accuracy, $a_1$ is one-shot accuracy, and $\alpha$ is coefficient.}
    \label{fig:context}
\end{figure}

The detailed parameters and performance of PFMs need to be considered in intelligent transportation systems of Metaverse are listed in Table I, including GPT3-13B~\cite{brown2020language}, GPT3-175B~\cite{brown2020language}, Uniformer-S~\cite{li2022uniformer}, Uniformer-B~\cite{li2022uniformer}, CLIP-ViT-L/14~\cite{cherti2022reproducible}, and CLIP-ViT-H/14~\cite{cherti2022reproducible}. As we can observe, only LFMs are large enough to have in-context learning ability, while  VFMs and MFMs are relatively small. As shown in Table I and Fig.~\ref{fig:context}, PFMs utilize meta-gradients to learn from context and improve performance as the user interacts with them. Then, the few-shot accuracy can be fit using the data in Table I. Therefore, contextual information has a corresponding impact on the quality of service provided by AIGC, such as the accuracy of PFMs. Although the introduction of context in PFMs can improve the model performance, the size of the context window also affects the resource consumption and latency during the inference of the model. As shown in Fig.~\ref{fig:accuracy}, the freshness and relevance of the examples in demonstrations decrease over time until it is no longer pertinent to the current generation task, which is rarely measured in previous work.

\subsection{Age of Context and Lease Context Algorithm}

Therefore, we propose the AoC for evaluating the relevance and freshness of examples in demonstrations that affect the quality of services of PFMs in currently executing downstream tasks. During inference of PFMs, the questions and answers can be recorded in the context windows as examples in demonstrations and instructions. These examples can be leveraged to improve the accuracy of PFMs as they can perform meta-gradient to fit these examples. However, the meta-gradient might have positive or negative effects on the accuracy, which depends on their quality, relevance, and timeliness. Similar to the age of information (AoI)~\cite{chen2021information}, the AoC indicates the relevance and timeliness of historical contextual examples in demonstrations to the cached PFM and the current inference task. As shown in Fig.~\ref{fig:accuracy}, the AoC follows the non-increasing age utility function, factoring with a vanishing coefficient of context. Based on the AoC, the number of examples in content can be calculated as the weighted sum of number of examples in demonstrations. Then, the accuracy of PFMs can be obtained by some function w.r.t. number of examples in context as the functions demonstrated in Fig.~\ref{fig:context}.

Finally, we introduce the LC algorithm, based on the AoC, to manage PFMs for mobile AIGC services efficiently. The LC algorithm tracks the number of examples in context, calculating them and removing the cached PFM with the least contexts, i.e., number of examples in context, when GPU memory is needed for loading a new PFM. This approach is effective for large numbers of PFMs on edge servers with limited GPU memory, prioritizing the removal of the least relevant PFM for the current inference task. Consequently, the accuracy of PFMs of mobile AIGC services is improved by leveraging more contextual information during inference.
\begin{table}[t]
\centering
\caption{Detailed system performance comparison for different caching algorithms.}
\label{tab:exp}
\begin{tabular}{c|ccccc}
\hline
              & Random & Cloud & FIFO  & LFU  & LC   \\ \hline
\makecell{System\\cost}          & 25.67  & 7.29  & 27.51 & 5.93 & \textbf{4.88} \\ \hline
   \makecell{Switching\\ cost}            & 18.72  & 0     & 23.28 & 0.37 & \textbf{0.32} \\ \hline
\makecell{Total \\accuracy\\cost}            & 0.13   & 0     & 0.52  & \textbf{0.36} &  0.44 \\ \hline
  \makecell{Average \\ accuracy\\cost}             & 0.0151   & 0     & 0.0085  & 0.0083 &  \textbf{0.0076} \\ \hline
\makecell{Inference \\latency}         &  0.12   & 0     & 1.30  & 1.32 & \textbf{1.26} \\ \hline
  \makecell{Offloading\\ latency} & 0.04   & 0     & 0.35  &  \textbf{0.24} & 0.31 \\ \hline
\makecell{Cloud \\cost}                & 6.63   & 7.29  &  \textbf{2.05}  & 3.63 & 2.52 \\ \hline
  \makecell{Edge \\Execution \\Ratio}    & 9.8\%   & 0\%     &  \textbf{70.7\%}  & 49.4\% & 65.0\%  \\ \hline
\end{tabular}
\end{table}

In the experiment, we compare the proposed LC algorithm with Random, cloud-only, first-in-first-out (FIFO), and least frequently used (LFU) baselines. With the objective of minimizing service latency and accuracy loss, the system cost is calculated as the sum of the switching cost, the total accuracy cost, the edge inference latency, the edge offloading latency, and the cloud computing cost. As listed in Table II, the performance of the proposed LC algorithm can achieve minimum total system cost while maintaining a high edge execution ratio, which indicates that most of the services are executed at edge servers. Especially, compared with the LFU algorithm, the least context (LC) algorithm can achieve a lower average service accuracy cost by efficiently leveraging the in-context learning ability of PFMs and contextual information.

% \section{Future Directions}\label{sec:future}

% \subsection{Model Compression and Memory-adaptive Models}

\section{Conclusions}\label{sec:conclusions}

In the article, we have studied edge caching and inference for serving PFMs in edge intelligence for Metaverse. We have proposed a joint model caching and inference framework for bringing the sparks of GPTs to mobile edge networks, toward achieving AGI. Specifically, we have proposed a new metric for evaluating the relevance and freshness of contextual examples and currently executing tasks. Furthermore, we have proposed the LC algorithm for cache replacement to improve the utilization of historical contextual information and thus increase the accuracy of mobile AIGC services. The experimental results demonstrate that the LC algorithm can reduce system costs and improve the execution ratio at edge servers.

\bibliographystyle{IEEEtran}
\bibliography{main}

\begin{IEEEbiographynophoto}{Minrui Xu} (minrui001@e.ntu.edu.sg)
    received the B.S. degree from Sun Yat-Sen University, Guangzhou, China, in 2021. He is currently working toward the Ph.D. degree in the School of Computer Science and Engineering, Nanyang Technological University, Singapore. His research interests mainly focus on mobile edge computing, deep reinforcement learning, and incentive mechanism design.
    \end{IEEEbiographynophoto}
    
    \begin{IEEEbiographynophoto}{Hongliang Zhang} [M’19] (hongliang.zhang92@gmail.com) is an assistant professor in the School of Electronics at Peking University, Beiijng, China. He was the recipient of the 2021 IEEE ComSoc
Heinrich Hertz Award.
    \end{IEEEbiographynophoto}

    \begin{IEEEbiographynophoto}{Dusit Niyato} [M'09, SM'15, F'17] (dniyato@ntu.edu.sg)
        is currently a professor in the School of Computer Science and Engineering, Nanyang Technological University, Singapore. He received the B.Eng. degree from King Mongkuts Institute of Technology Ladkrabang (KMITL), Thailand in 1999 and Ph.D. in electrical and computer engineering from the University of Manitoba, Canada in 2008. His research interests are in the areas of Internet of Things (IoT), machine learning, and incentive mechanism design.
    \end{IEEEbiographynophoto}
    \begin{IEEEbiographynophoto}{Jiawen Kang} [M'18]
  received the Ph.D. degree from the Guangdong University of Technology, China in 2018. He was a postdoc at Nanyang Technological University, Singapore from 2018 to 2021. He currently is a  professor at Guangdong University of Technology, China. His research interests mainly focus on blockchain, security, and privacy protection in wireless communications and networking.
    \end{IEEEbiographynophoto}

    \begin{IEEEbiographynophoto}{Zehui Xiong} [M'20]
        (zehui\_xiong@sutd.edu.sg) is an Assistant Professor at Singapore University of Technology and Design. Prior to that, he was a researcher with Alibaba-NTU Joint Research Institute, Singapore. He received the Ph.D. degree in Computer Science and Engineering at Nanyang Technological University, Singapore. He was a visiting scholar with Princeton University and University of Waterloo. His research interests include wireless communications, network games and economics, blockchain, and edge intelligence.
    \end{IEEEbiographynophoto}

\begin{IEEEbiographynophoto}{Shiwen Mao}  [S’99, M’04, SM’09, F'19] 
        (smao@ieee.org) received his
Ph.D. in electrical and computer engineering from Polytechnic University, Brooklyn, NY. He is a Professor and Earle C. Williams Eminent Scholar, and Director of the Wireless Engineering Research and Education Center at Auburn University. His
research interests include wireless networks and multimedia communications.
    \end{IEEEbiographynophoto}

   \begin{IEEEbiographynophoto}{Zhu Han} [S’01, M’04, SM’09, F’14] (zhuhan22@gmail.com) currently is a professor in the Electrical and Computer Engineering Department at the University of Houston, Texas. He has been an AAAS Fellow since 2019. He received the IEEE Kiyo Tomiyasu Award in 2020. He has been a 1 percent highly cited researcher since 2017 according to Web of Science.
    \end{IEEEbiographynophoto}
\end{document}